\documentclass{jfm}
\usepackage{graphicx}
\usepackage{epstopdf, epsfig}
\usepackage{color}
\usepackage{soul}
\usepackage{graphicx}
\usepackage{float}
\usepackage{overpic}
\usepackage{amsmath}
\usepackage{amssymb}
\usepackage{caption}
\usepackage{overpic}
\usepackage{multirow}

\newcommand{\Vp}{\tilde{\bf{u}}}

\newcommand{\hatu}{\hat{\bf{u}}}

\shorttitle{Global instability of laminar separation bubbles}
\shortauthor{W. He}
\title{Global instability of long separation bubbles in a laminar boundary layer}
\author{Wei  He\aff{}
  \corresp{\email{whe@liverpool.ac.uk}}}

\affiliation{School of Engineering, University of Liverpool, The Quadrangle, Brownlow Hill, Liverpool L69 3GH, United Kingdom
}

\begin{document}

\maketitle

\begin{abstract}
This work aims to numerically investigate the linear global instability of long separation bubbles origin from the changes in the adverse pressure gradient inside a laminar flat plate  boundary layer disturbed by placing a bluff body with small clearances. The boundary layer starts at $Re_{\delta^\ast}$=121.7 and is steady in the computational extent after being disturbed by a NACA 4415 airfoil and a cylinder with clearance $h\leqslant 0.2$, respectively. It is found that the dominant stability is associated with  two- and three-dimensional stationary eigenmodes.  There is a  less damped oscillation mode at a small wavenumbers range. The strong  effect of three-dimensionality  is further confirmed in a  non-modal analysis framework. Transient growth  analysis shows that the most strongest optimal perturbation is three-dimensional in the range of parameters studied.  
The modal analysis reveals that the instability grows with reducing the clearance, while the optimal perturbations grow with increasing the clearance.  It is found that the separated boundary layer stability analysis is less dependent on the bluff body geometries.

\end{abstract}


\section{Introduction}
The mechanisms of transition induced by  a ``short'' separation bubble  in a laminar boundary flow is well studied by \cite{spalart_strelets_2000}, in which Tollmien-Schlichting and Kelvin-Helmholtz modes are addressed. 
The instability of a  laminar separation bubble generated by varying the top gradient in a flat-plate boundary layer flow is investigated by  \cite{vassilis2000rsta}, and the three-dimensional stationary mode is firstly reported.
 \citet{marquillie2003jfm} analysed the instability of flat plate boundary flow accompanied with a long separation bubble arising after a small bump. The low frequency (less than 0.1) vortex shedding has been revealed for the flow higher that the critical Reynolds number. Similar  three-dimensional  flow stability is investigated by \cite{gallaire2007jfm}. The  transition mechanism of short laminar separation bubbles with turbulent reattachment  is addressed by \cite{alam2000jfm} at high Reynolds numbers. The Tollmien-Schlichting wave behavior in flat plate boundary layer disturbed by a small hump or micro-vortex generators are numerous numerically investigated \citep{xu2016, jp2017tcfd}, which kind of mechanism is further verified in experiments \citep{ye2016jfm,puckertRist2018jfm}. The topology characteristics of laminar separation bubbles in flat plate boundary layer flow are studied by global linear instability theory \citep{RodriguezTheofilis2010jfm}. \cite{rist2002} identified two classifies of instability existing  outside the boundary layer and near the wall in a flat plate boundary layer flow.

The linear instability analysis for the spanwise homogeneous  NACA airfoils in steady open flow are addressed in the literature  \citep{TheofilisSherwin,he2017jfm}. They found that the traveling Kelvin-Helmholtz (KH) mode is the dominant mechanisms in the wake instability. 
While taking account of ground effect in the flow past a NACA 4415 airfoil with a high angle of attack  $AoA=20^\circ$ at low Reynolds numbers, \cite{he2018ast} analysed the instability of the massively separated flow through imposing an analytical Blasius velocity profile as the {\it inlet} boundary condition accompanied with a stationary ground, which result a thick boundary layer comparable to the chord. The primary instability of the stable flow in the range of clearance $h\in(0.2,\infty)$ at $Re=500-1000$ (based on chord) showed that the KH instability dominating the two- and three-dimensional instabilities. When the clearance reduced to $h=0.2$ in a steady flow $Re=500$, the airfoil is already immersed in the boundary layer, a thin and long laminar separation occurs in the downstream along the flat ground.  Overall, the flow can not sustain fully attached along the ground anymore, and it becomes a more complex boundary layer flow with two separation bubbles coexisting inside it. One short separation stays behind the airfoil and another longer one is adhere to the flat ground downstream. The global  stationary instability mode  grows and is trying to overcome the KH mode \citep{he2018ast}.

When the separation behind the airfoil  coupled into the boundary layer, a disturbance will be added into the boundary layer.  To continue reducing the clearance, the long separation bubble would increase and  may further  expedite the downstream transition. This is unlike the laminar separation arisen by a varying velocity function on the top of the computational domain \citep{vassilis2000rsta}.  In the current investigated flow configuration, it is unknown which stability mode dominates the flow. 
In order to further investigate this phenomenon  and explore the potential stability mechanisms, global linear stability of a steady flat plate laminar boundary layer flow disturbed by two kind of geometry bluff bodies  near the plate with small clearance $h\leqslant 0.2$  is systematically studied in this paper.

\section{Problem formation}
The original flat plate boundary layer flow under the effect of adverse pressure gradient is disturbed by adding a bluff body: a NACA 4415 airfoil holding a  high angle of attack 20 degrees and  a circular cylinder, respectively. Direct numerical simulation is employed to investigate the instabilities on account of ground  effect between bluff body and flat plate.
A rectangular computational domain is defined as  $\{x,y\}\in \{[-10, 25c]\times[0,15c]\}$ and  is dicretized into $O(2000)$  quadrilateral elements using the software $Gmsh$ \citep{gmsh}, where $c$ is the nondimensional airfoil chord. The NACA 4415 airfoil is placed above a non-inclined flat plate with clearance $h=0.1$ and $0.2$, respectively, where the clearance $h$ is defined as the vertical distance between trailing edge (its $x$-coordinate is fixed at $x=1.0$)  of the airfoil and the flat plate. In order to release the effect of geometry, an alternative cylinder is used for comparison. The cylinder is right-side aligned with  the airfoil and has the same transverse projection area as the airfoil.  An analytical Blasius velocity profile  is imposed at {\it inlet}  boundary ($x=-10$). The dimensionless Reynolds number $Re_{\delta^\ast}\equiv\frac{U_\infty \delta^\ast}{\nu}=121.7$ is defined basing on the flat plate boundary layer displacement thickness $\delta^\ast$ at inlet, where  $U_\infty$ is the freestream velocity, and $\nu$ is the fluid kinematic viscosity. 
~The flat plate and airfoil or cylinder are set to be {\it no-slip} boundary conditions, the rest boundaries are set to be {\it outflow}. 

The base flow is governed  by the two-dimensional incompressible Navier-Stokes   and continuity equations
\begin{eqnarray}
{\partial_t {\bf u} }+ {\bf u} \bcdot \nabla {\bf u} &= - \nabla p +{Re_c^{-1}}\nabla^2 {\bf u},\qquad
\nabla\bcdot\bf{u} &=0.
\label{eqn:nse}
\end{eqnarray}
Where the $\bf u$ and $p$ are the dimensionless velocity vector and pressure of the fluid, $Re_c$ is the Reynolds number based on the airfoil chord length $c$ and freestream velocity $U_\infty$. According to the kinematic viscosity $\nu$, $Re_c$ is also fixed as  500 in current study. Following  the biglobal linear instability theory \citep{TheofilisARFM},  the total field can be decomposed into a steady state (or time-periodic) flow and a small-amplitude perturbation $\Vp(x,y,z)=\hatu (x,y) e^{i(\beta z-\lambda t)}$. Here, $\lambda=\lambda_r+i\lambda_i$ is the complex eigenvalue of matrix $\mathsfbi L$,  $\beta$ is the real spanwise wavenumber.  Then the equation \ref{eqn:nse} can be linearized into an eigenvalue system
\begin{eqnarray}
{\mathsfbi L}\hatu = -i\lambda \hatu.
\label{eqn:eig}
\end{eqnarray}
If there is at least one mode, such that $\lambda_i >0$, then the flow is said to be unstable. In the same time, the  evolution of initial perturbation in the boundary layer governed by the equation \ref{eqn:eig} in a short time interval $\tau$ is examined using the non-modal theory as addressed by \cite{luchini2000}.  For a given initial value $\Vp_0$,  its maximum energy gain is defined as $G(\tau)=\frac{<{\mathsfbi L}^+(\tau) {\mathsfbi L}(\tau) \Vp_0, \Vp_0>}{<\Vp_0,\Vp_0>}$, where $\mathsfbi L^+$ is the adjoint operator of $\mathsfbi L$. The value of $G(\tau)$ can be decided by solving the dominant eigenvalue of the singular value decomposition problem  \citep{Schmid2012}.
In this paper, both base flow and stability analysis are performed  and cross-validated using two open-source spectral/hp element method codes {\it Nektar++} \citep{Cantwell2015205} and  {\it Semtex} \citep{BlackburnSherwin2004}. The boundary conditions setup for stability analysis can be referred to  \cite{he2018ast}, as well as the mesh resolution and  convergence studies. 

\section{Results}
\subsection{Base flow}
\label{sec:baseflow}

According to the current computational extent, the maximum $Re_x$ is $1\times10^4$ at the end of  domain for the plat plate boundary layer flow dominated by the adverse pressure gradient. Here ``$x$'' is the distance measured from nominal leading edge of the flat plat. This value is much less than the critical transition number $Re_x=3.5\times10^5$ \citep[p.~453]{Schlichting} ensuring the flow is steady, laminar,  and no separation in the entire extent. After putting a bluff body, such as cylinder or  airfoil at high angle of attack in the boundary layer, the adverse pressure gradient in downstream will be changed, which may result a  separation. Previous study shows that the flow past a NACA 4415 airfoil with $AoA=20^\circ$ is steady with the clearance $h\leqslant 0.3$ \citep{he2018ast}, and there is no separation closed to the flat plate when $h=0.3$, except a large recirculation zone behind the airfoil. 

\begin{figure}
 \centering
\begin{overpic}[width=0.925 \textwidth]{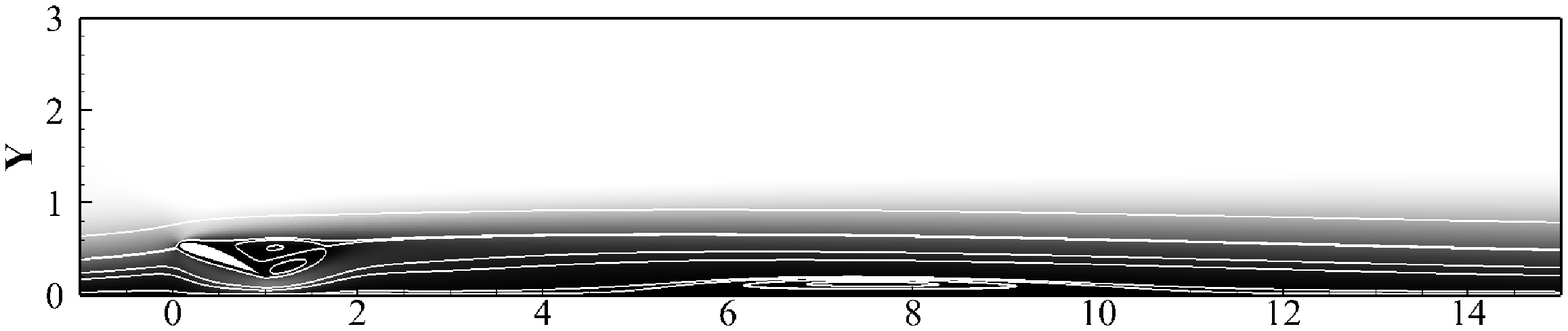}\put(0, 20) {($a$)} \end{overpic}
\begin{overpic}[width=0.925 \textwidth]{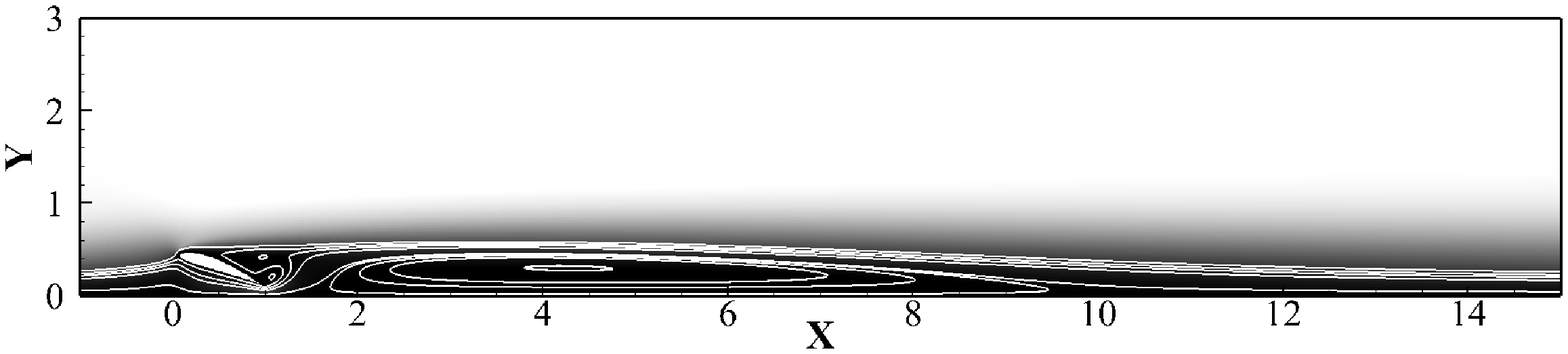}\put(0, 20) {($b$)} \end{overpic}
 \caption{Streamlines of the base flow shows the long recirculation zone appears by placing the airfoil near the flat plate with clearances: ($a$): $h=0.2$, ($b$): $h=0.1$. Gray scale visualizes the streamwise velocity $u$.}
 \label{fig:longbubble}
 \end{figure}

\begin{figure}
 \centering
 \includegraphics[width=0.925 \textwidth]{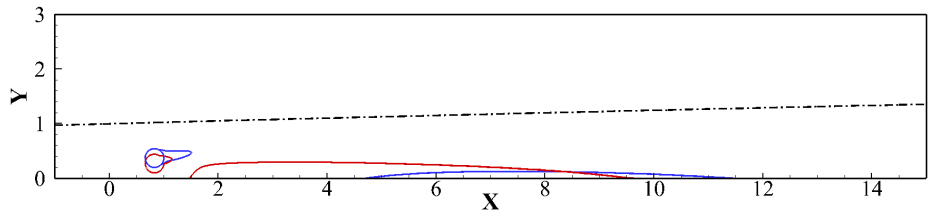}
 \caption{Separation bubble comparison induced by placing cylinders at different clearance values. The black dashed-dotted line represent the boundary boarder of the flat plate flow.  Blue and red lines represent the reversed flow boarders corresponding to $h=0.2$ and 0.1, respectively.}
 \label{fig:separation_cyl}
 \end{figure}

When reducing the clearance $h$ to 0.2, another long and thin laminar separation bubble appears in  downstream region. Continually reducing  $h$ to 0.1, except the separation bubble immediately after the airfoil  is shrunk, there is a long  and thick separation bubble growing in the boundary layer closed to the flat plate, which covers the range of $x\in(1.6, 10)$, as shown in figure \ref{fig:longbubble}. Replacing the airfoil with a cylinder, which diameter equals to the transverse projection length of the airfoil, the flow still keeps in a steady state, and it has a  similar  thin and thick long separation bubbles in the boundary layer with clearance $h=0.2$ and 0.1, respectively (see figure \ref{fig:separation_cyl}). However, opposite to  the flow with the airfoil  where a recirculation zone behind it, there is no recirculation behind the cylinder with the same clearance  $h=0.1$. Kindly reminder, the recirculation zone is more thicker than the boarder $u=0$ (denoted by red/blue color lines) determined, but this will not change the separation and reattachment points of the bubble. When the clearance increased, the recirculation zone grows behind the cylinder, but the separation bubble adheres to the flat plate decreased.

\begin{figure}
 \begin{center}
 \includegraphics[width=0.925 \textwidth]{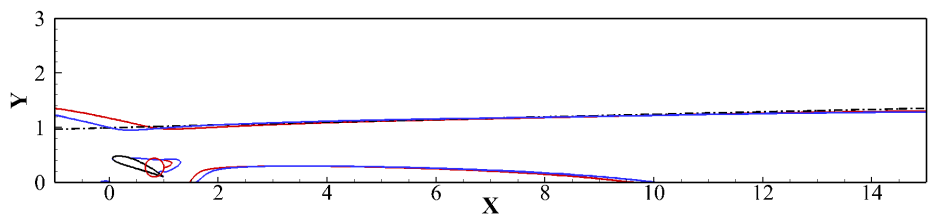}
 \caption{Separation bubble comparison induced by cylinder and airfoil  with the clearance $h=0.1$, respectively. The boundary layer boarders are added.}
 \label{fig:separation_cyl_foil_h0o1}
 \end{center}
 \end{figure}

\begin{figure}
 \begin{center}
 \includegraphics[width=0.925 \textwidth]{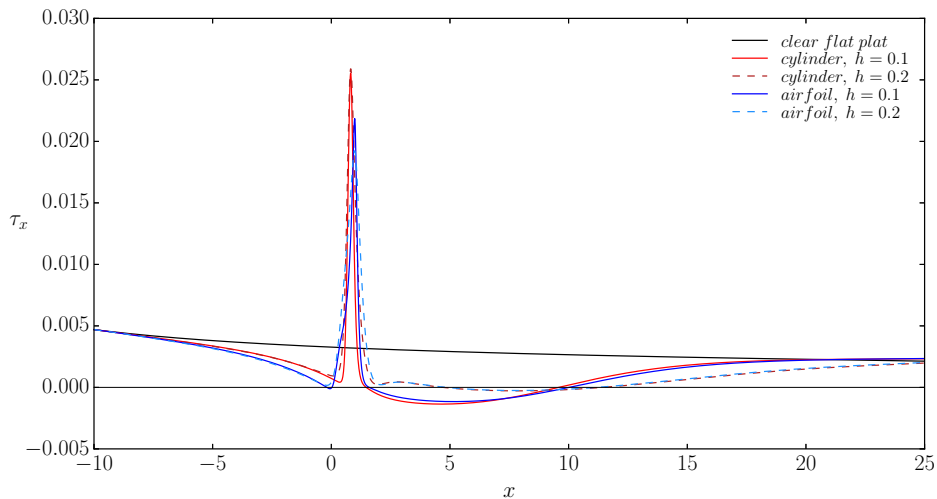}
 \caption{Comparison of wall shear stress $\tau_x$ along the flat plat.}
 \label{fig:wallstress}
 \end{center}
 \end{figure}

Figure \ref{fig:separation_cyl_foil_h0o1} compares the long separation bubbles produced by airfoil and cylinder at $h=0.1$. Basically, there is no significant difference between two flows: the long bubbles dominate nearly a large same size  domain, as listed in Table \ref{tab:sep-reattach}.  Owing to the flow past the lowest point of cylinder is earlier than  airfoil's trailing edge, the position of the separation bubble is slightly earlier. Further more, the boundary layer thickness (the flat boundary layer boarder is outlined by the black dash-dotted line) does not change compared to the flat plate flow in the downstream after the flow past one unit distance of the bodies.  More completely comparison by plotting the wall shear stress $\tau_x$ of the plate is shown in figure \ref{fig:wallstress}. The wall shear stress of the flat plate boundary layer flow has a gradually decreased  parabolic distribution (outlined by the thick black solid line). For the flow past the cylinder or airfoil, the wall shear stress of flat plate drops closed to zero firstly, then it abruptly increases to its maxima  around $x=1.0$, which values are about 5 times larger than on the flat plate. Those four peaks cover the range of the bluff bodies. Since there is a long separation bubble, the wall shear stress has a negative value, which corresponding to the separation point ($SP$) and reattachment point ($RP$) of the separation bubble ( more details are listed in Table \ref{tab:sep-reattach}). After the flow reattached, the wall shear stress is  recovered gradually. 
Table \ref{tab:sep-reattach} also shows the reverse flow level $u_{res}=||u_{min}/u_{max}||$ in the long separation bubbles. The maximum $u_{res}$ is about 10\%, which is less than 15\% to excite an absolute instability in a boundary layer flow \citep{alam2000jfm}.  

\begin{table}
\begin{center}
\begin{tabular}{crrr|rrr}
       &\multicolumn{3}{c}{airfoil} &  \multicolumn{3}{c}{cylinder}  \\\hline
 $h$    & $SP$  & $RP$  &  $u_{res}(\%)$  &  $SP$  & $RP$  &  $u_{res}(\%)$  \\\hline
0.1     & 1.60  & 10.02 & 9.96	& 1.46 & 9.74 & 4.59\\
0.2     & 4.80  & 11.28 & 9.58	& 4.67 & 11.45 & 6.56 
\end{tabular}
\caption{\label{tab:sep-reattach}Separation ({\it lef column}) and reattachment points ({\it right column})  of the separation bubble along the flat plate.}
\end{center}
\end{table}

\subsection{Primary instability analysis}
\label{sec:modal}

The long time behavior of perturbations evolution can be examined by solving equation \ref{eqn:eig}. The temporal  stability analysis is carried out  following the scheme reviewed by \citet{TheofilisARFM}.  In the stability analysis of a NACA airfoil post-stall flow at low Reynolds numbers, the KH stability dominates the two- and three-dimensional flow at small wavenumbers $\beta$, and stationary mode decays slower than KH mode in the large wavenumber region \citep{ GIORIA201588, he2017jfm}.  When a NACA 4415 airfoil flow is confined by  an infinite long and nonslip flat plate under it with a clearance $h\geqslant 0.3$ at $Re_c=500$, KH  instability still plays a dominant role in the flow system \citep{he2018ast}.  

\begin{figure} 
 \centering
\begin{overpic}[width=0.95\textwidth]{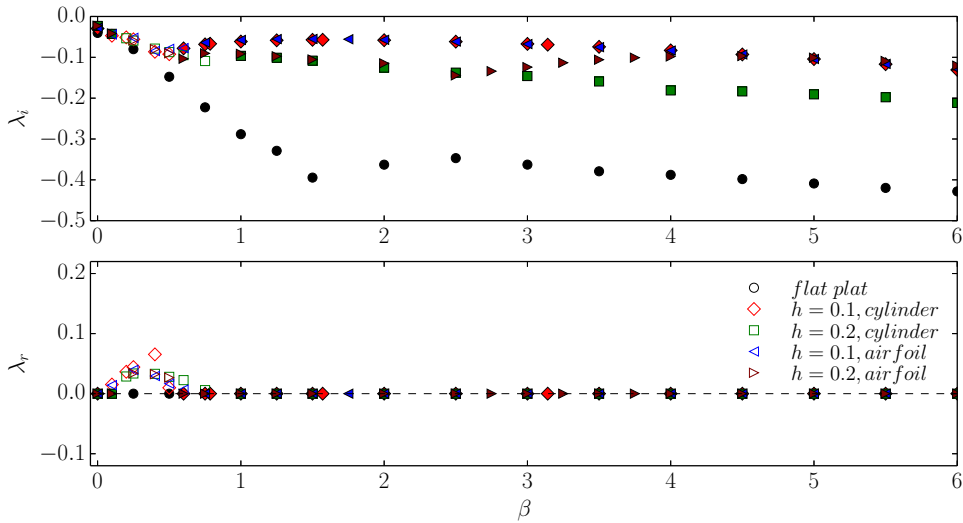}
\end{overpic}
 \caption{Growth rate ($\lambda_i$) and frequency ($\lambda_r$) of the leading eigenvalues  comparison by placing cylinder and airfoil at different clearances $h$. Solid symbols are Stationary modes and open symbols represent oscillation modes.}
 \label{fig:modal_cyl_foil_h0o1}
 \end{figure}

In the clean flat plate  boundary layer flow at $Re_{\delta^{\ast}}$=121.7,  the flow is naturally steady, and the two-dimensional stationary  mode is most amplified.  In a  range of wavenumbers ($0< \beta < 6$) examined in this paper, the three-dimensional stability is dominated by the stationary mode, as represented by the dark black circles  in figure \ref{fig:modal_cyl_foil_h0o1}, from it we can find that no oscillation mode exists in this flow.


As shown in figure \ref{fig:modal_cyl_foil_h0o1}, the stability becomes more unstable when a long separation bubble  appearing in the boundary layer than the flat plate flow.
 The three-dimensional stationary stability dominates the spanwise wavenumber at $\beta>1$ with the clearance $h=0.1$.  In a small range ($0<\beta<1$), there is a low frequency ($\lambda_r<0.1$) oscillation mode becoming more amplified, which is consistent with  the found in a bump disturbed unsteady boundary layer flow \citep{marquillie2003jfm}.
When $h=0.2$, the stability in both cylinder and airfoil disturbed flow are consistent with each other till to $\beta=2.5$. After that, the growth rate ($\lambda_i$) of cylinder disturbed continues decaying, while in another case, $\lambda_i$ turns off to the level of $h=0.1$. Overall, the growth rate in all cases are less than zero which indicates that  the flows are asymptotic stable. 

\begin{figure} 
 \centering
\begin{overpic}[width=0.925 \textwidth]{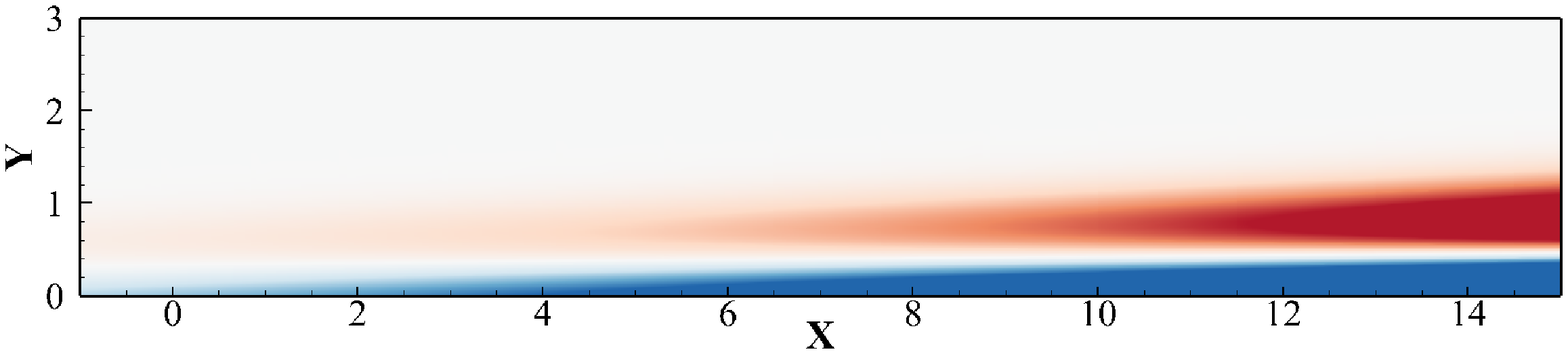} \put(0, 20) {($a$)} \end{overpic}
 \begin{overpic}[width=0.925 \textwidth]{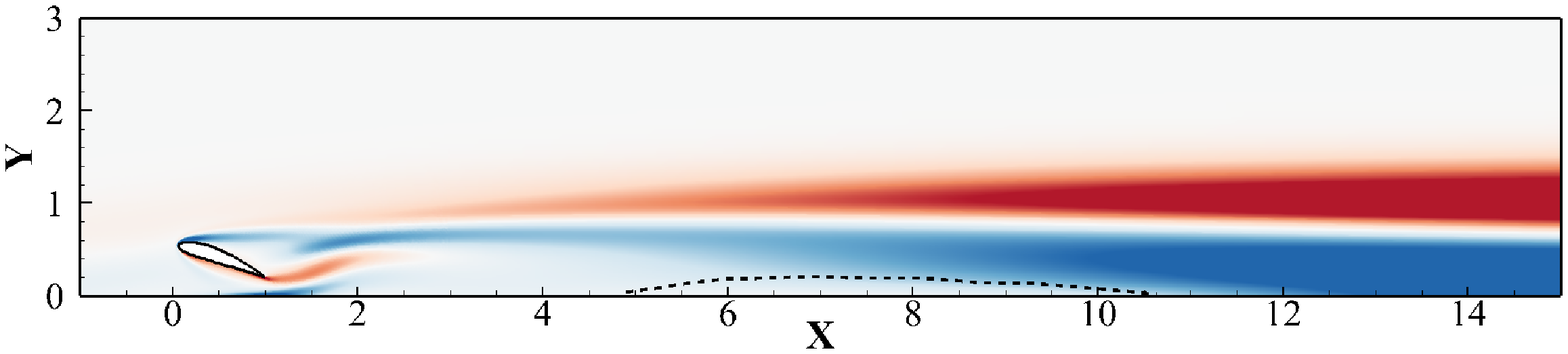} \put(0, 20) {($b$)} \end{overpic}
 \begin{overpic}[width=0.925 \textwidth]{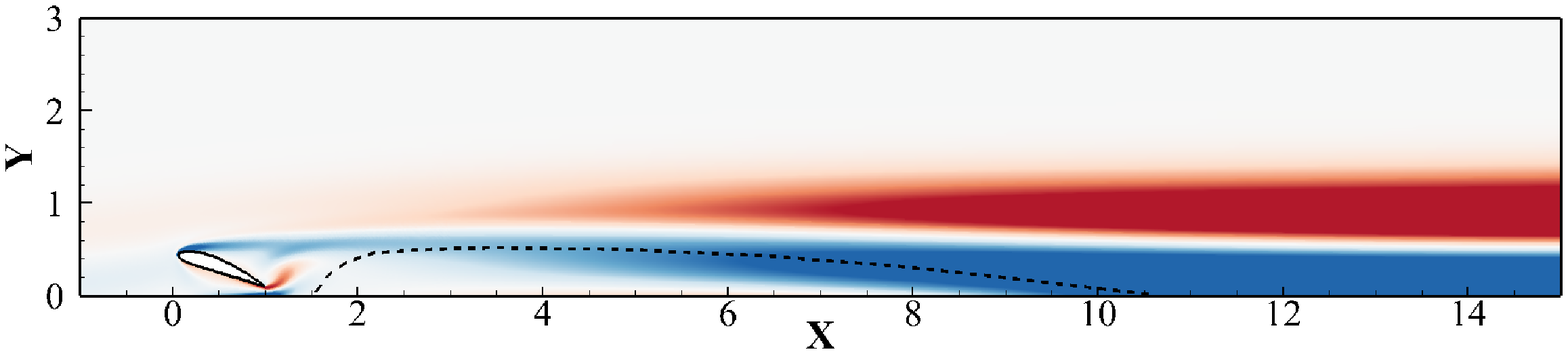} \put(0, 20) {($c$)} \end{overpic}
 \caption{Normalized spanwise vorticity amplitude function $\hat{\omega}_z$  of the least stable mode in the  clear flat plat flow ($a$), disturbed by a airfoil with a larger clearance $h=0.2$ ($b$) and  and with a smaller clearance $h=0.1$ ($c$), respectively. The blue/red color represents $\hat{\omega}_z \in [-0.1, 0.1]$. Black dashed lines denote the thin- and thick-long separation bubbles in corresponding base flows. }
 \label{fig:modal_amplitude}
 \end{figure}

\begin{figure} 
 \centering
\includegraphics[width=0.725\textwidth]{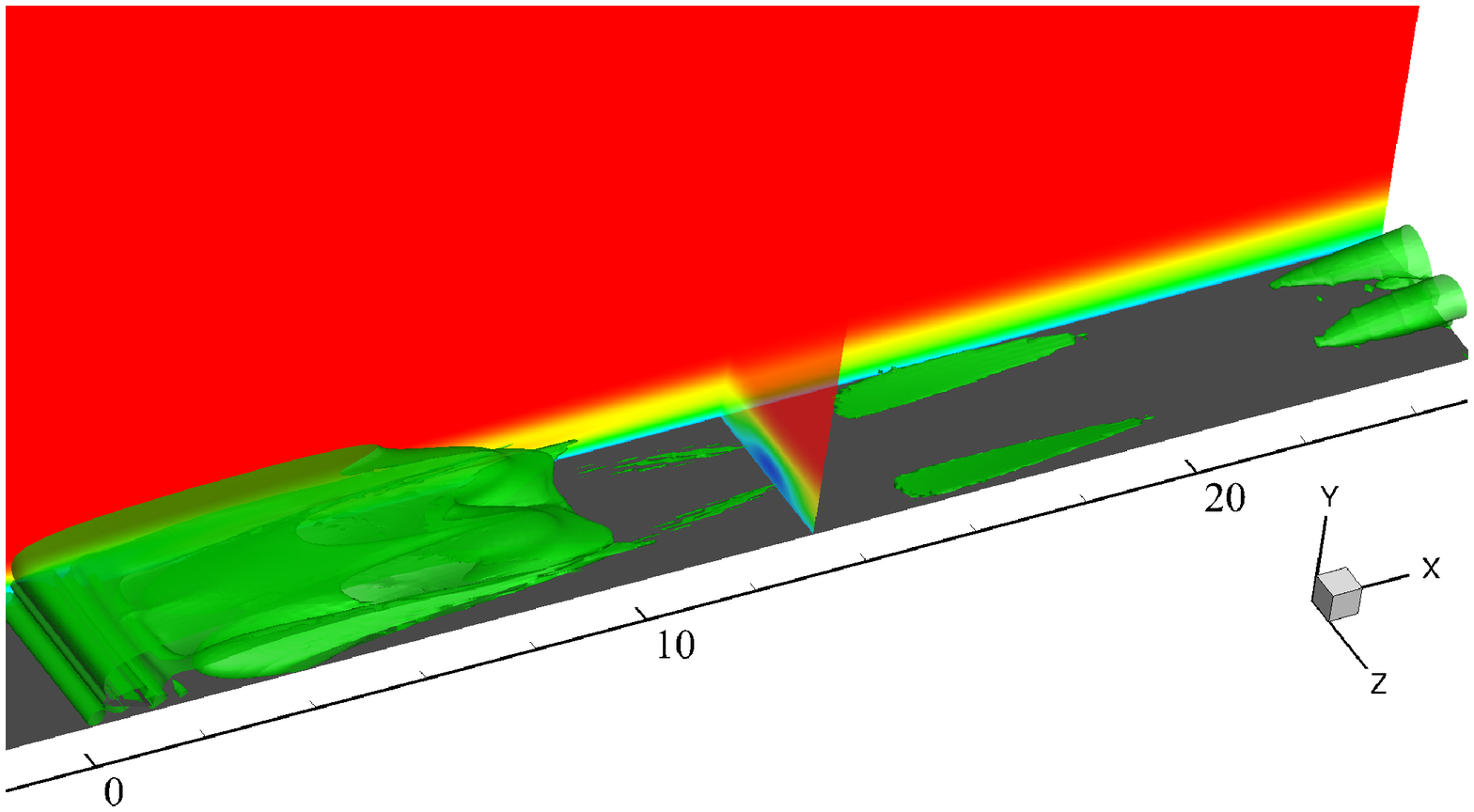} 
 \caption{Three-dimensional reconstruction of superimposing a stationary mode  of $\beta=2$ onto the long and thick laminar separation base flow at a level of $\epsilon=2\times10^{-4}$. The vortical structure is represented by $\lambda_2$-criterion (green), the flat plat is visualized by gray color, and the rest parts plot the streamwise velocity.}
 \label{fig:3d}
 \end{figure}

Figure \ref{fig:modal_amplitude} shows the comparison of the normalized amplitude function of spanwise vorticity $\hat{\omega}_z$ of the least stable   mode at $\beta=0$ for clear flat plate boundary layer flow and the cases containing long and short separation bubbles generated by airfoil. In one side, the stability grows with the developing of boundary layer (figure \ref{fig:modal_amplitude} ($a$)), in another side, there is a minor growth in the shear layer near the leading- and trailing-edge of the airfoil. Unlike the found in \citep{marquet2009}, the stability inside the separation bubble  is much weaker than that outside, as seen in figure \ref{fig:modal_amplitude} ($b$) and ($c$). 
The three-dimensional of flow construction is shown in figure \ref{fig:3d}. A stationary mode ($\beta=2$) is linearly superimposed on the long and thick laminar separation bubble generated by the airfoil at $h=0.1$, using a level of $\epsilon=2\times10^{-4}$ of the freestream magnitude.  The vortical structure is firstly shown by a kind of complicated separation bubble till around $x=10$, then followed by a pair of oblique streak vortex mode (about 3 chord length) adhering to the plat. A pair of more parallel vortex grows downstream and finally converted into roll-up vortices.

\subsection{Non-modal analysis}
\label{sec:nonmodal}
\begin{figure}
 \centering
\begin{overpic}[width=0.48 \textwidth]{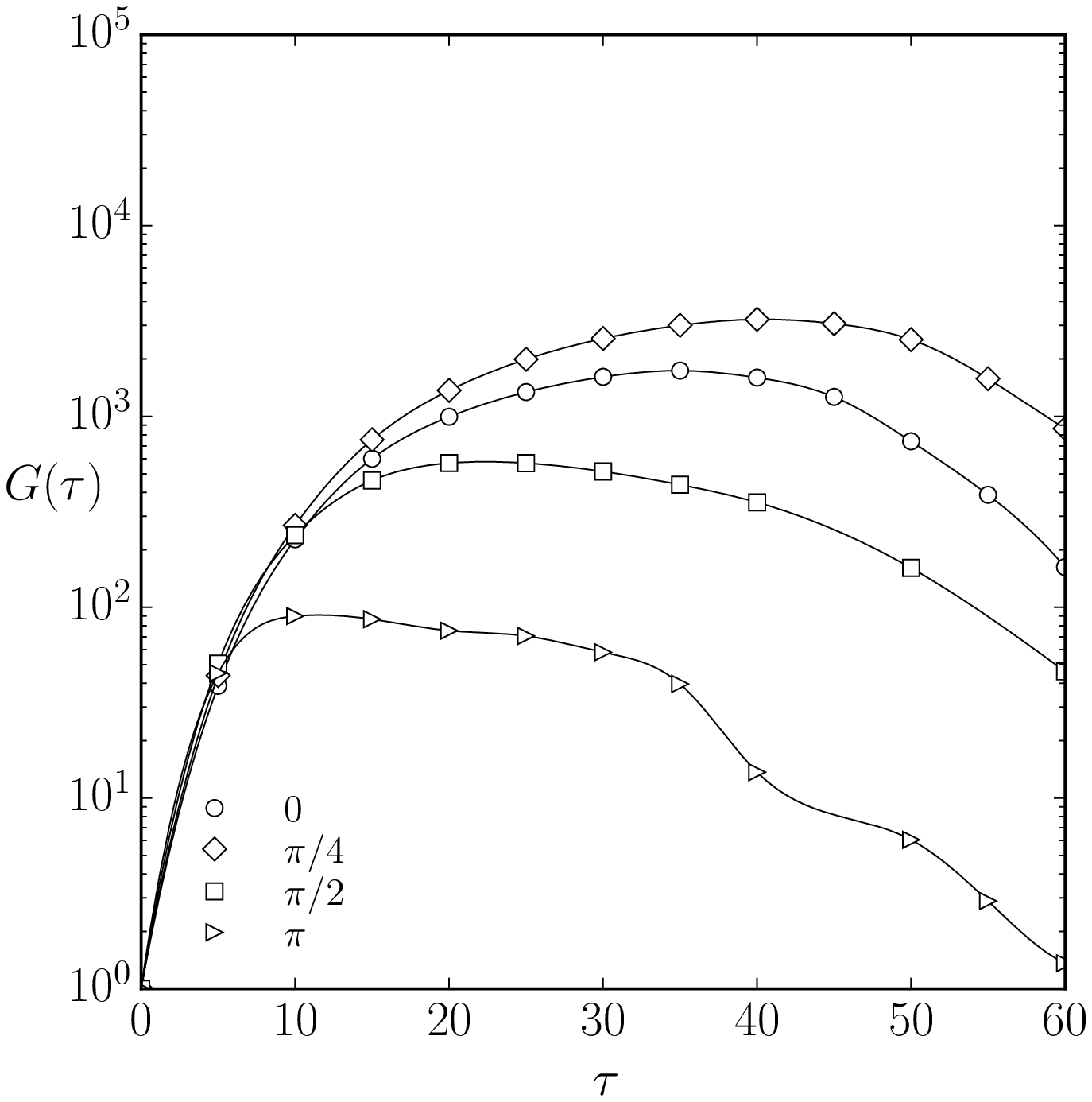}\put(0, 91) {($a$)} \end{overpic}\quad
\begin{overpic}[width=0.48 \textwidth]{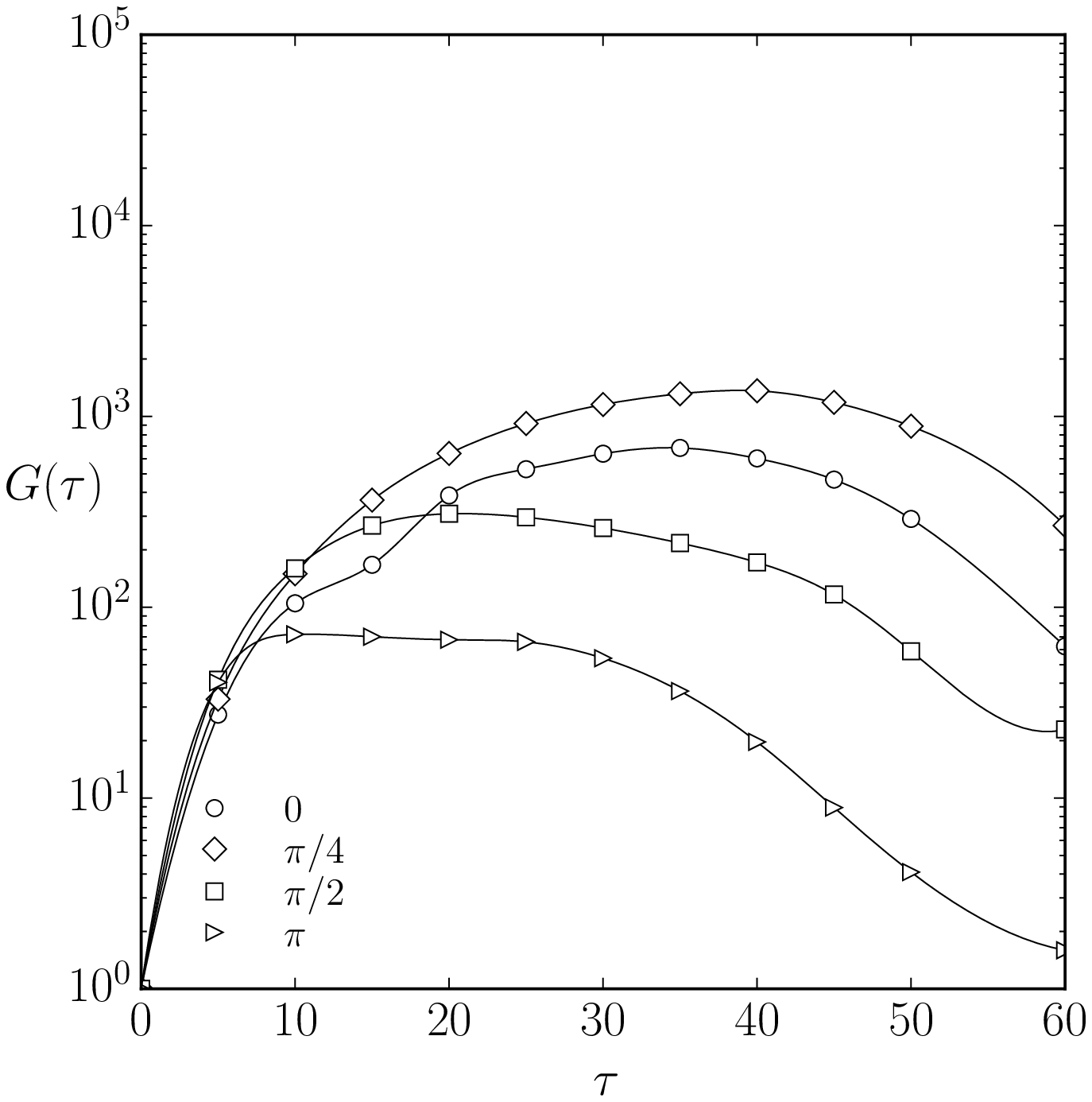}\put(0, 91) {($b$)} \end{overpic}
 \caption{Optimal gain $G(\tau)$ as a function of wavenumber $\beta$ in the  boundary layer disturbed by ($a$): NACA 4415 airfoil, ($b$): cylinder with clearance $h=0.1$.}
 \label{fig:tg_foil_h0o1}
 \end{figure}

The optimal perturbations are investigated using the scheme of non-modal analysis or {\it transient growth}, as addressed by \cite{Nadir2009} in the flow past circular cylinder as well as NACA airfoils flow \citep{GIORIA201588, he2017jfm}. Previous studies show that the dominant optimal perturbation in the flow past a NACA 0015 airfoil at high Reynolds number with a small angle of attack is two-dimensional \citep{LohBlackburnSherwin}. Analogous conclusion is also found for three NACA airfoils at low Reynolds numbers open flow containing a large recirculation zone \citep{he2017jfm}. 
Transient growth is investigated in the range of spanwise wavenumber $0\leqslant \beta \leqslant \pi$ based on the steady base flow presented in section \ref{sec:baseflow}. 
In figure \ref{fig:tg_foil_h0o1}($a$), the flow configuration has a long and thick separation downstream and a clearance $h=0.1$ between the airfoil and the plate. The gain $G(\tau)$ of the optimal perturbation  in short time horizons $\tau < 10$ has no significant difference. While, as the integral time  $\tau > 10$, the two-dimensional optimal gain is overcome by the three-dimensional perturbation at $\beta=\frac{\pi}{4}$.  
As $\beta \geqslant \frac{\pi}{2}$, the gain grows slower than that in two-dimensional flow. The optimal perturbation for cylinder case at $h=0.1$ has the same behavior as the  airfoil (see figure \ref{fig:tg_foil_h0o1}($b$)), but its maxima value is smaller compared to the former flow.  

\begin{figure} 
 \centering
\begin{overpic}[width=0.48 \textwidth]{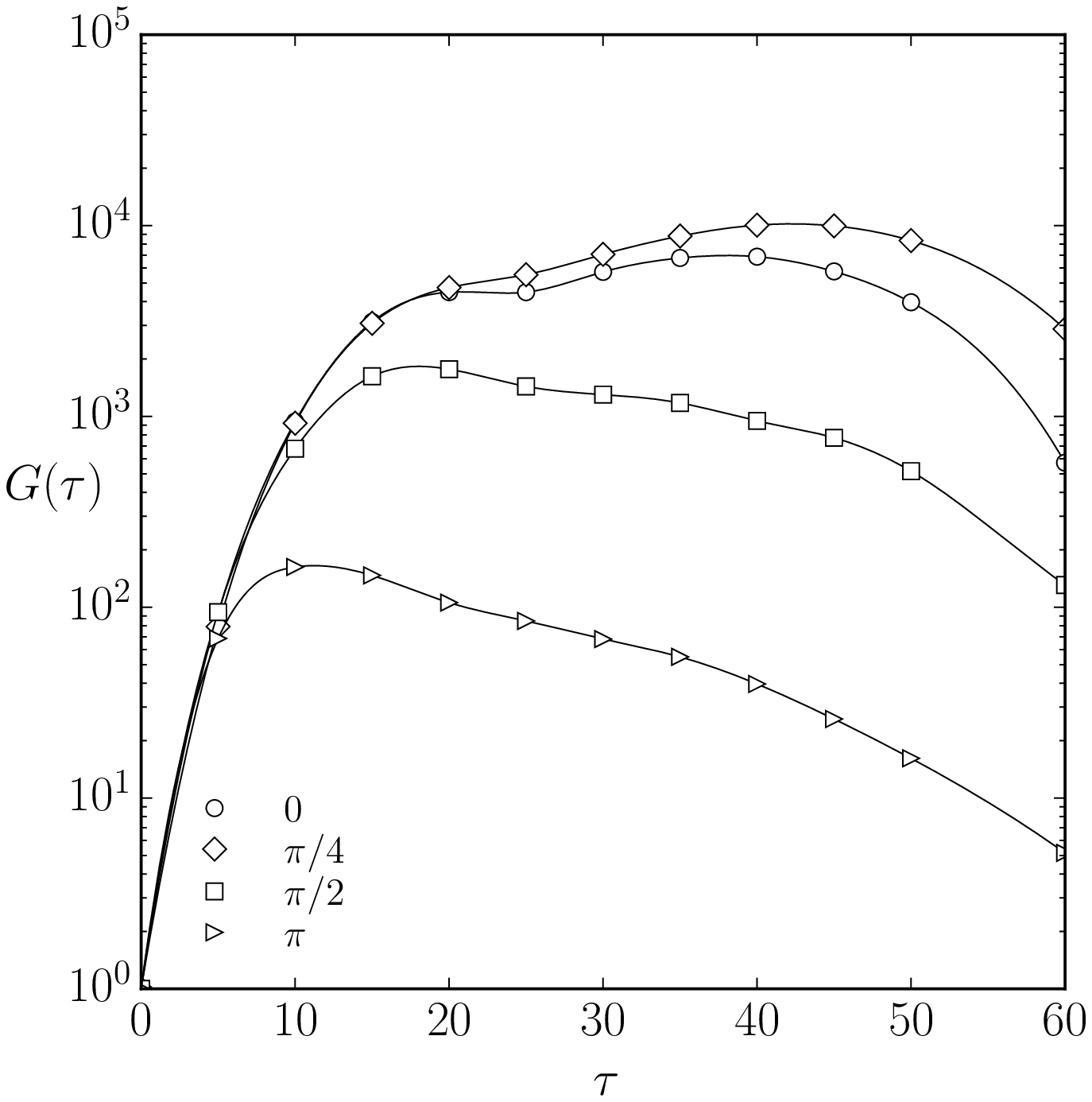}\put(0, 91) {($a$)} \end{overpic}\quad
\begin{overpic}[width=0.48 \textwidth]{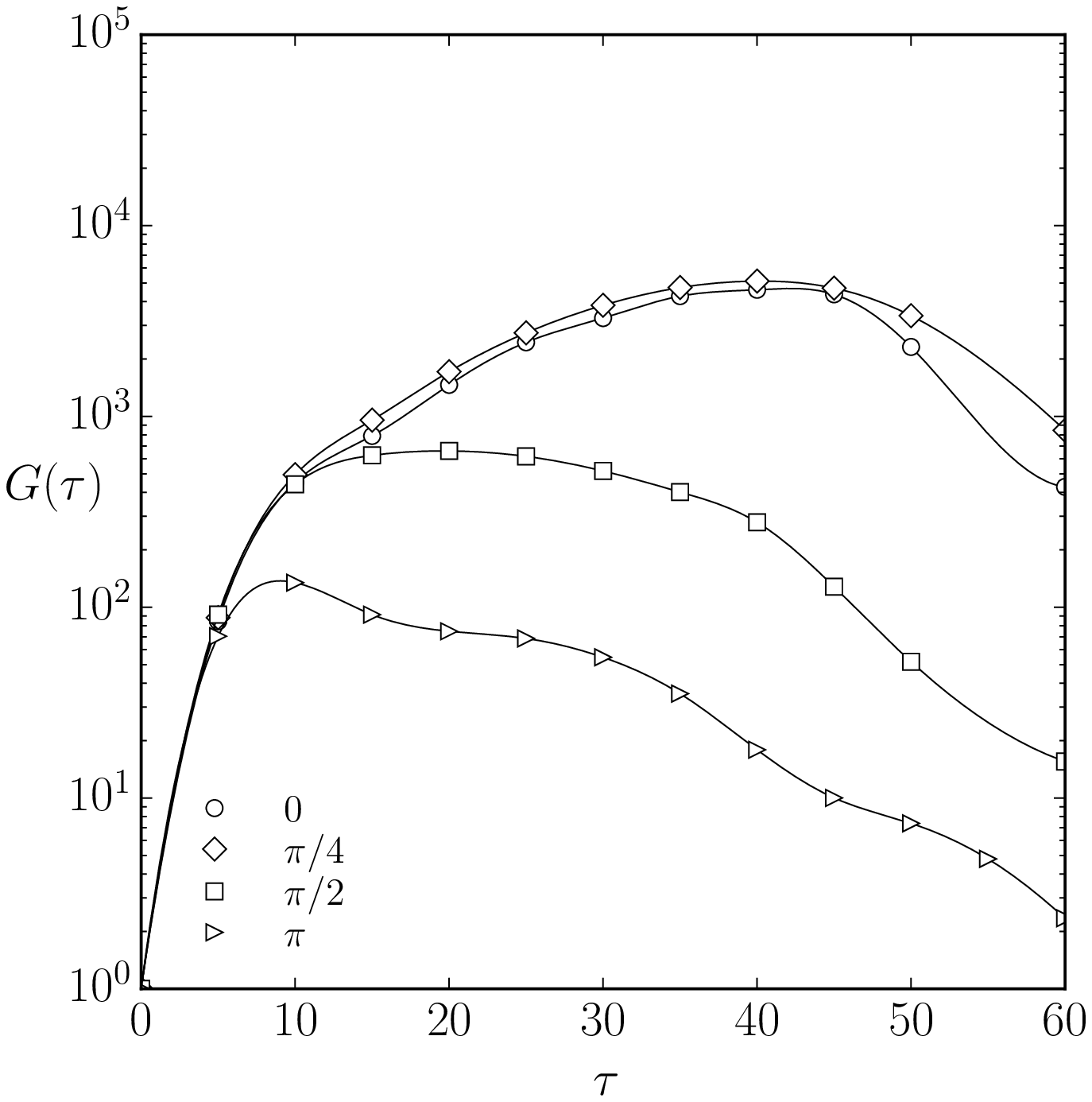}\put(0, 91) {($b$)} \end{overpic}
  \caption{Optimal gain $G(\tau)$ as a function of wavenumber $\beta$ in the  boundary layer disturbed by ($a$): NACA 4415 airfoil, ($b$): cylinder with clearance $h=0.2$.}
 \label{fig:tg_foil_h0o2}
 \end{figure}

Figure \ref{fig:tg_foil_h0o2}  shows the  transient growth in the boundary layer at different wavenumbers with clearance $h=0.2$ containing  the airfoil and the cylinder, respectively. Three-dimensional optimal perturbations still play a dominant role. Quantitatively, the energy increased in the range of parameters examined with larger clearance. This has maybe resulted by a large recirculation zone behind the bluff bodies, and the long and thick separation bubble contributes less in the short time transient growth of the boundary layer, which in contrast, has more contribution to the modal instability. 
At asymptotic long times, the  gain obtained in the non-modal analysis reveals the initial optimal perturbations evolve into KH modes and  are damped quickly compared to the analysis  in NACA airfoils \citep{he2017jfm} and cylinder flows \citep{Nadir2009}.

 \section{Conclusions}
The instability analysis of a laminar flat plate boundary layer flow with and without disturbed by bluff bodies near the plate is performed. When a NACA airfoil stands  just above the  flat plate with clearance 0.2 chord at high angle of attack, a thin and long separation bubble appears in the downstream. While reducing the clearance, a  more thick and long separation bubble appears and moves backwards to the airfoil. Similar phenomenon can be observed by replacing the airfoil with a cylinder. The long separation bubbles are less dependent on the bluff body geometries. Modal analysis reveals that the stability in these flows are dominated by three-dimensional stationary mode, and there is a low frequency traveling mode existing in a small wavenumber range. The instability grows with reducing the clearance between bluff bodies and the flat plate. Non-modal analysis reveals that the three-dimensional optimal perturbations play a more important role than the two-dimensional one. The optimal perturbation gain grows with increasing the clearance, but it starts to be overtaken by  the  two-dimensional optimal perturbations. 
\bibliographystyle{jfm}
\bibliography{blsep}

\begin{thebibliography}{25}
\expandafter\ifx\csname natexlab\endcsname\relax\def\natexlab#1{#1}\fi
\def\au#1{#1} \def\ed#1{#1} \def\yr#1{#1}\def\at#1{#1}\def\jt#1{\textit{#1}}
  \def\bt#1{#1}\def\bvol#1{\textbf{#1}} \def\vol#1{#1} \def\pg#1{#1}
  \def\publ#1{#1}\def\arxiv#1{#1}\def\org#1{#1}\def\st#1{\textit{#1}}

\bibitem[Abdessemed {\em et~al.\/}(2009)Abdessemed, Sharma, Sherwin \&
  Theofilis]{Nadir2009}
{\sc \au{Abdessemed, N.}, \au{Sharma, A.~S.}, \au{Sherwin, S.~J.} \&
  \au{Theofilis, V.}} \yr{2009}  \at{Transient growth analysis of the flow past
  a circular cylinder}.  \jt{Phys.~Fluids}  \bvol{21}~(4),  \pg{044103}.

\bibitem[Alam \& Sandham(2000)]{alam2000jfm}
{\sc \au{Alam, M.} \& \au{Sandham, N.D.}} \yr{2000}  \at{Direct numerical
  simulation of {`short'} laminar separation bubbles with turbulence
  reattachment}.  \jt{J.~Fluid Mech.}  \bvol{410},  \pg{1--28}.

\bibitem[Blackburn \& Sherwin(2004)]{BlackburnSherwin2004}
{\sc \au{Blackburn, H.~M.} \& \au{Sherwin, S.~J.}} \yr{2004}  \at{Formulation
  of a galerkin spectral element--fourier method for three--dimensional
  incompressible flows in cylindrical geometries}.  \jt{J. Comput. Phys.}
  \bvol{197}~(2),  \pg{759--778}.

\bibitem[Cantwell {\em et~al.\/}(2015)Cantwell, Moxey, Comerford, Bolis, Rocco,
  Mengaldo, Grazia, Yakovlev, Lombard, Ekelschot, Jordi, Xu, Mohamied,
  Eskilsson, Nelson, Vos, Biotto, Kirby \& Sherwin]{Cantwell2015205}
{\sc \au{Cantwell, C.D.}, \au{Moxey, D.}, \au{Comerford, A.}, \au{Bolis, A.},
  \au{Rocco, G.}, \au{Mengaldo, G.}, \au{Grazia, D.~De}, \au{Yakovlev, S.},
  \au{Lombard, J.E.}, \au{Ekelschot, D.}, \au{Jordi, B.}, \au{Xu, H.},
  \au{Mohamied, Y.}, \au{Eskilsson, C.}, \au{Nelson, B.}, \au{Vos, P.},
  \au{Biotto, C.}, \au{Kirby, R.M.} \& \au{Sherwin, S.J.}} \yr{2015}
  \at{Nektar++: An open-source spectral/hp element framework}.
  \jt{Comput.~Phys.~Commun.}  \bvol{192},  \pg{205--219}.

\bibitem[Gallaire {\em et~al.\/}(2007)Gallaire, Marquillie \&
  Ehrenstein]{gallaire2007jfm}
{\sc \au{Gallaire, F.}, \au{Marquillie, M.} \& \au{Ehrenstein, U.}} \yr{2007}
  \at{{Three-dimensional} transverse instabilities in detached boundary
  layers}.  \jt{J.~Fluid Mech.}  \bvol{571},  \pg{221--233}.

\bibitem[Geuzaine \& Remacle(2009)]{gmsh}
{\sc \au{Geuzaine, C.} \& \au{Remacle, J.~F.}} \yr{2009}  \at{Gmsh: a
  three-dimensional finite element mesh generator with built-in pre- and
  post-processing facilities}.  \jt{Int.~J.~Numer.~Meth.~Eng.}  \bvol{79}~(11),
   \pg{1309--1331}.

\bibitem[Gioria {\em et~al.\/}(2015)Gioria, He \& Theofilis]{GIORIA201588}
{\sc \au{Gioria, R.~S.}, \au{He, W.} \& \au{Theofilis, V.}} \yr{2015}  \at{On
  global linear instability mechanisms of flow around airfoils at low reynolds
  number and high angle of attack}.  \jt{Procedia IUTAM}  \bvol{14},
  \pg{88--95}.

\bibitem[He {\em et~al.\/}(2017)He, Gioria, P\'erez \& Theofilis]{he2017jfm}
{\sc \au{He, W.}, \au{Gioria, R.~S.}, \au{P\'erez, J.~M.} \& \au{Theofilis,
  V.}} \yr{2017}  \at{Linear instability of low {Reynolds} number massively
  separated flow around three {NACA} airfoils}.  \jt{J.~Fluid Mech.}
  \bvol{811},  \pg{701--741}.

\bibitem[He {\em et~al.\/}(2018)He, Yu \& Li]{he2018ast}
{\sc \au{He, W.}, \au{Yu, P.} \& \au{Li, L.K.B.}} \yr{2018}  \at{Ground effect
  on the stability of separated flow around a {NACA} 4415 airfoil at low
  {Reynolds} numbers}.  \jt{Aerosp. Sci. Technol.}  \bvol{72},  \pg{64--82}.

\bibitem[Loh {\em et~al.\/}(2014)Loh, Blackburn \&
  Sherwin]{LohBlackburnSherwin}
{\sc \au{Loh, S.~A.}, \au{Blackburn, H.~M.} \& \au{Sherwin, S.~J.}} \yr{2014}
  Transient growth in an airfoil separation bubble.  \bt{In {\em 19th
  Australasian Fluid Mechanics Conference\/}}. Melbourne, Australia, Dec 8-11,
  2014.

\bibitem[Luchini(2000)]{luchini2000}
{\sc \au{Luchini, P.}} \yr{2000}  \at{Reynolds-number-independent instability
  of the boundary layer over a flat surface: optimal perturbations}.  \jt{J.
  Fluid Mech.}  \bvol{404},  \pg{289--309}.

\bibitem[Marquet {\em et~al.\/}(2009)Marquet, Lombardi, Chomaz, Sipp \&
  Jacquin]{marquet2009}
{\sc \au{Marquet, O.}, \au{Lombardi, M.}, \au{Chomaz, J.-M.}, \au{Sipp, D.} \&
  \au{Jacquin, L.}} \yr{2009}  \at{Direct and adjoint global modes of a
  recirculation bubble: lift-up and convective non-normalities}.  \jt{J. Fluid
  Mech.}  \bvol{622},  \pg{1–21}.

\bibitem[Marquillie \& Ehrenstein(2003)]{marquillie2003jfm}
{\sc \au{Marquillie, M.} \& \au{Ehrenstein, U.}} \yr{2003}  \at{On the onset of
  nonlinear oscillations in a separating {boundary-layer} flow}.  \jt{J.~Fluid
  Mech.}  \bvol{490},  \pg{169--188}.

\bibitem[Mart{\'i}n \& Paredes(2017)]{jp2017tcfd}
{\sc \au{Mart{\'i}n, J.~A.} \& \au{Paredes, P.}} \yr{2017}
  \at{Three-dimensional instability analysis of boundary layers perturbed by
  streamwise vortices}.  \jt{Theor. Comput. Fluid Dyn.}  \bvol{31}~(5),
  \pg{505--517}.

\bibitem[Puckert \& Rist(2018)]{puckertRist2018jfm}
{\sc \au{Puckert, Dominik~K.} \& \au{Rist, Ulrich}} \yr{2018}  \at{Experiments
  on critical reynolds number and global instability in roughness-induced
  laminar--turbulent transition}.  \jt{J. Fluid Mech.}  \bvol{844},
  \pg{878--904}.

\bibitem[Rist \& Maucher(2002)]{rist2002}
{\sc \au{Rist, U.} \& \au{Maucher, U.}} \yr{2002}  \at{Investigations of
  time--growing instabilities in laminar separation bubbles}.  \jt{Eur.
  J.~Mech.~(B/Fluids)}  \bvol{21},  \pg{495--509}.

\bibitem[Rodr\'iguez \& Theofilis(2010)]{RodriguezTheofilis2010jfm}
{\sc \au{Rodr\'iguez, D.} \& \au{Theofilis, V.}} \yr{2010}  \at{Structural
  changes of laminar separation bubbles induced by global linear instability}.
  \jt{J.~Fluid Mech.}  \bvol{655},  \pg{280--305}.

\bibitem[Schlichting(1979)]{Schlichting}
{\sc \au{Schlichting, H.}} \yr{1979} {\em Boundary Layer Theory\/}, 7th edn.
  \publ{McGraw-Hill}.

\bibitem[Schmid \& Henningson(2012)]{Schmid2012}
{\sc \au{Schmid, P.} \& \au{Henningson, D.}} \yr{2012} {\em Stability and
  Transition in Shear Flows\/}.  \publ{Springer--Verlag New York}.

\bibitem[Spalart \& Strelets(2000)]{spalart_strelets_2000}
{\sc \au{Spalart, P.~R.} \& \au{Strelets, M.~KH.}} \yr{2000}  \at{Mechanisms of
  transition and heat transfer in a separation bubble}.  \jt{J. Fluid Mech.}
  \bvol{403},  \pg{329--349}.

\bibitem[Theofilis(2011)]{TheofilisARFM}
{\sc \au{Theofilis, V.}} \yr{2011}  \at{Global linear instability}.  \jt{Annu.
  Rev. Fluid Mech.}  \bvol{43},  \pg{319--352}.

\bibitem[Theofilis {\em et~al.\/}(2000)Theofilis, Hein \&
  Dallmann]{vassilis2000rsta}
{\sc \au{Theofilis, V.}, \au{Hein, S.} \& \au{Dallmann, U.}} \yr{2000}  \at{On
  the origins of unsteadiness and {three-dimensionality} in a laminar
  separation bubble}.  \jt{Phil.~Trans.~R.~Soc.~Lond.~A}  \bvol{358},
  \pg{3229--3246}.

\bibitem[Theofilis \& Sherwin(2001)]{TheofilisSherwin}
{\sc \au{Theofilis, V.} \& \au{Sherwin, S.~J.}} \yr{2001} Global instabilities
  in trailing- edge laminar separated flow on a {NACA} 0012 aerofoil.  \bt{In
  {\em Proceedings of the XV International Symposium on Airbreathing Engines
  ISABE 2001-1094\/}}. Bangalore, India, September 3-7, 2001.

\bibitem[Xu {\em et~al.\/}(2016)Xu, Sherwin, Hall \& Wu]{xu2016}
{\sc \au{Xu, H}, \au{Sherwin, SJ}, \au{Hall, P} \& \au{Wu, X}} \yr{2016}
  \at{The behaviour of tollmien-schlichting waves undergoing small-scale
  localised distortions}.  \jt{J. Fluid Mech.}  \bvol{792},  \pg{499--525}.

\bibitem[Ye {\em et~al.\/}(2016)Ye, Schrijer \& Scarano]{ye2016jfm}
{\sc \au{Ye, Q.}, \au{Schrijer, F.} \& \au{Scarano, F.}} \yr{2016}
  \at{Boundary layer transition mechanisms behind a micro-ramp}.  \jt{J. Fluid
  Mech.}  \bvol{793},  \pg{132--161}.

\end{thebibliography}


\begin{thebibliography}{14}
\expandafter\ifx\csname natexlab\endcsname\relax\def\natexlab#1{#1}\fi

\bibitem[Batchelor(1971)]{Batchelor59}
{\sc Batchelor, G.~K.} 1971 Small-scale variation of convected quantities like
  temperature in turbulent fluid. part 1. general discussion and the case of
  small conductivity. {\em J.~Fluid Mech.\/} {\bf 5}, 113--133.

\bibitem[Brownell \& Su(2004)]{Brownell04}
{\sc Brownell, C.~J. \& Su, L.~K.} 2004 Planar measurements of differential
  diffusion in turbulent jets. {\em AIAA Paper 2004-2335\/}.

\bibitem[Brownell \& Su(2007)]{Brownell07}
{\sc Brownell, C.~J. \& Su, L.~K.} 2007 Scale relations and spatial spectra in
  a differentially diffusing jet. {\em AIAA Paper 2007-1314\/}.

\bibitem[Dennis(1985)]{Dennis85}
{\sc Dennis, S. C.~R.} 1985 {Compact explicit finite difference approximations
  to the Navier--Stokes equation}. In {\em Ninth Intl Conf. on Numerical
  Methods in Fluid Dynamics\/} (ed. Soubbaramayer \& J.~P. Boujot), {\em
  Lecture Notes in Physics\/}, vol. 218, pp. 23--51. Springer.

\bibitem[Hwang \& Tuck(1970)]{Hwang70}
{\sc Hwang, L.-S. \& Tuck, E.~O.} 1970 On the oscillations of harbours of
  arbitrary shape. {\em J.~Fluid Mech.\/} {\bf 42}, 447--464.

\bibitem[Koch(1983)]{Koch83}
{\sc Koch, W.} 1983 Resonant acoustic frequencies of flat plate cascades. {\em
  J.~Sound Vib.\/} {\bf 88}, 233--242.

\bibitem[Lee(1971)]{Lee71}
{\sc Lee, J.-J.} 1971 Wave-induced oscillations in harbours of arbitrary
  geometry. {\em J.~Fluid Mech.\/} {\bf 45}, 375--394.

\bibitem[Linton \& Evans(1992)]{Linton92}
{\sc Linton, C.~M. \& Evans, D.~V.} 1992 The radiation and scattering of
  surface waves by a vertical circular cylinder in a channel. {\em Phil.\
  Trans.\ R. Soc.\ Lond.\/} {\bf 338}, 325--357.

\bibitem[Martin(1980)]{Martin80}
{\sc Martin, P.~A.} 1980 On the null-field equations for the exterior problems
  of acoustics. {\em Q.~J. Mech.\ Appl.\ Maths\/} {\bf 33}, 385--396.

\bibitem[Miller(1991)]{Miller91}
{\sc Miller, P.~L.} 1991 Mixing in high schmidt number turbulent jets. PhD
  thesis, California Institute of Technology.

\bibitem[Rogallo(1981)]{Rogallo81}
{\sc Rogallo, R.~S.} 1981 Numerical experiments in homogeneous turbulence. {\em
  Tech. Rep.\/} 81835. NASA Tech.\ Mem.

\bibitem[Ursell(1950)]{Ursell50}
{\sc Ursell, F.} 1950 Surface waves on deep water in the presence of a
  submerged cylinder i. {\em Proc.\ Camb.\ Phil.\ Soc.\/} {\bf 46}, 141--152.

\bibitem[{van Wijngaarden}(1968)]{Wijngaarden68}
{\sc {van Wijngaarden}, L.} 1968 On the oscillations near and at resonance in
  open pipes. {\em J.~Engng Maths\/} {\bf 2}, 225--240.

\bibitem[Worster(1992)]{Worster92}
{\sc Worster, M.~G.} 1992 {The dynamics of mushy layers}. In {\em In
  Interactive dynamics of convection and solidification\/} (ed. S.~H. Davis,
  H.~E. Huppert, W.~Muller \& M.~G. Worster), pp. 113--138. Kluwer.

\end{thebibliography}

\end{document}